# How LinkedIn Economic Graph Bonds Information and Product: Applications in LinkedIn Salary

Xi Chen, Yiqun Liu, Liang Zhang, Krishnaram Kenthapadi
LinkedIn Corporation, USA
(xchen1, yiqliu, lizhang, kkenthapadi)@linkedin.com

## ABSTRACT

The LinkedIn Salary product was launched in late 2016 with the goal of providing insights on compensation distribution to job seekers, so that they can make more informed decisions when discovering and assessing career opportunities. The compensation insights are provided based on data collected from LinkedIn members and aggregated in a privacy-preserving manner. Given the simultaneous desire for computing robust, reliable insights and for having insights to satisfy as many job seekers as possible, a key challenge is to reliably infer the insights at the company level when there is limited or no data at all. We propose a two-step framework that utilizes a novel, semantic representation of companies (*Company2vec*) and a Bayesian statistical model to address this problem. Our approach makes use of the rich information present in the LinkedIn Economic Graph, and in particular, uses the intuition that two companies are likely to be similar if employees are very likely to transition from one company to the other and vice versa. We compute embeddings for companies by analyzing the LinkedIn members' company transition data using machine learning algorithms, then compute pairwise similarities between companies based on these embeddings, and finally incorporate company similarities in the form of peer company groups as part of the proposed Bayesian statistical model to predict insights at the company level. We perform extensive validation using several different evaluation techniques, and show that we can significantly increase the coverage of insights while, in fact, even slightly improving the quality of the obtained insights. For example, we were able to compute salary insights for 35 times as many title-region-company combinations in the U.S. as compared to previous work, corresponding to 4.9 times as many monthly active users. Finally, we highlight the lessons learned from practical deployment of our system.

## 1. INTRODUCTION

Online professional social networks and job platforms such as LinkedIn play a key role in ensuring an efficient labor marketplace, by connecting talent (job seekers) with opportunity (job providers). Considering that salary is known to be an important factor when looking for new opportunities [2, 3], products such as LinkedIn Salary have the potential to reduce asymmetry of compensation knowledge, and to serve as market-perfecting tools for job seekers and job providers [15].

The LinkedIn Salary product, launched in November 2016, allows members to explore compensation insights by searching for different titles and regions. For each (title, region) combination, we present the distribution of base salary, bonus, and other types of compensation, the variation of pay based on factors such as experience, education, company size, and industry, and the highest paying regions, industries, and companies. These insights are generated based on data collected from LinkedIn members, using a combination of techniques such as encryption, access control, de-identification, aggregation, and thresholding for preserving privacy of users [19], and modeling techniques such as outlier detection and Bayesian hierarchical smoothing for ensuring robust, reliable insights [18].

A key challenge in this application is the simultaneous need for ensuring sufficient product coverage (having insights to satisfy as many job seekers as possible) and computing robust, reliable compensation insights. At the time of launch, the product only allowed LinkedIn members to discover compensation insights by searching for a title and a region, and then exploring other dimensions. However, we noticed that a large number of LinkedIn members were interested in learning about compensation insights at the company level, as reflected from their feedback. Consequently, there was a strong desire to generate compensation insights for as many (title, region, company) *cohorts* as possible, and to make such insight pages available as part of the product user experience. While previous work [18] used statistical modeling techniques to compute robust insights, a crucial limitation was that the insights were provided only for cohorts with at least a few member submissions, and in particular, the existing system could provide insights for only about $35K$ (title, region, company) cohorts, covering a small fraction of LinkedIn's monthly active users. Such low coverage caused poor user experience, making it impossible to include company-level insight pages as part of the product.

We address the problem of reliably inferring compensation insights at the company level, that is, *predicting insights for (title, region, company) cohorts with no member-submitted data at all*. The intuition underlying our approach is that two companies can be considered similar if employees are very likely to transition from one company to the other and vice versa. In the context of computing compensation insights, this assumption is rooted in the observation that job transitions typically result in higher pay: in a study of over 5000 job moves, 63% resulted in same or higher base pay, with 2.1% average pay raise for those who moved to a different company [22].

Our solution mines the rich information present in the LinkedIn Economic Graph [26] to generate a novel, semantic representation

(embedding) of companies. We propose an algorithm for learning company embeddings from the LinkedIn members' company transition data (*Company2vec*), compute pairwise similarity values between companies based on these embeddings, and then define the peer company group for each company as the set of most similar companies. Finally, we incorporate company similarities as part of a proposed Bayesian statistical model to predict insights at the company level, wherein we combine the estimates for (title, region) component and company adjustment.

We demonstrate the efficacy of our models through extensive validation with de-identified compensation data collected from several million LinkedIn members, and show that we can significantly increase the coverage of insights while, in fact, even slightly improving the quality of the obtained insights. As an example, our techniques enable the computation of base salary insights for 35 times as many (title, region, company) combinations in the U.S. as compared to previous work [18], corresponding to a coverage for 4.9 times as many monthly active users. Finally, we present the lessons learned in practice from the deployment of our system.

## 2. PROBLEM SETTING

The LinkedIn Salary product enables users to explore compensation insights (e.g., percentiles and histograms) for different titles, locations, and companies, as in Figure 1. The shown insights are based on the compensation data that we have been collecting from corresponding LinkedIn members (using a give-to-get model), which are then processed using techniques such as encryption, access control, de-identification, thresholding, aggregation, and outlier detection to ensure member privacy and data quality [19, 18]. Due to privacy requirements, the salary modeling system can only access cohort level data containing de-identified compensation submissions (e.g., compensation entries for Software Engineers working at LinkedIn in San Francisco Bay Area), limited to those cohorts that contain at least a minimum number of entries. Since the empirical percentile estimates are not reliable for those cohorts with very little data, the existing system used a Bayesian hierarchical smoothing methodology, which exploited the hierarchical structure amongst the cohorts and "borrowed strength" from the ancestral cohorts to derive estimates for such small-sized cohorts. First, the ancestral cohort that can "best explain" the observed entries in the given cohort is chosen as the "best ancestor." The data from this ancestral cohort is used as the *prior*, and the *posterior* of the cohort of interest is obtained based on the *prior* and the observed entries [18]. However, this methodology was designed only for cohorts with member submitted entries, and in particular, cannot be used to provide reliable insights for (title, region, company) cohorts with no data at all, for which there is tremendous interest in the product. This issue becomes more acute since the number of (title, region, company) cohorts with member provided data is quite small, covering a small fraction of LinkedIn's monthly active users, and hence, increasing the product coverage (both in terms of the number of cohorts with reliable insights and the corresponding covered fraction of monthly active users) is a critical business requirement.

To address these twin business goals, we have built a modeling system consisting of two components: (1) Computation of pairwise company similarity and peer company groups based on LinkedIn's economic graph data containing company transitions by LinkedIn members, and (2) Inference of compensation insights for (title, region, company) cohorts with no data using a Bayesian statistical model, that utilizes the company similarity and peer company group information (§3). We have integrated this system as part of the existing LinkedIn Salary modeling architecture (§4), and also created a standalone interface for other LinkedIn applications to access / benefit from these insights as well as the peer company group information (e.g., for improved job seniority filtering in the job recommendation application [8]). We next highlight the key modeling challenges that are addressed by our framework.

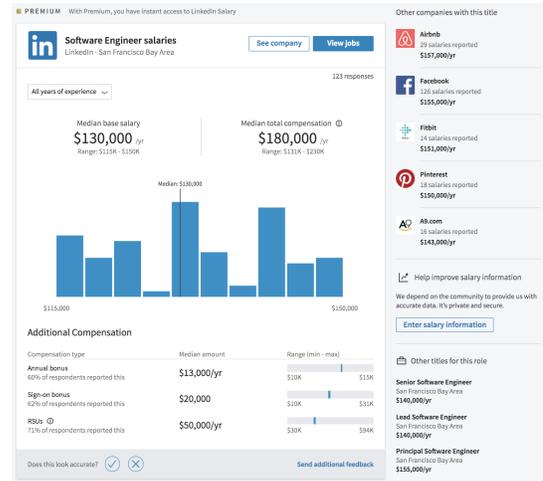

Figure 1: LinkedIn Salary insight page for (Software Engineer, San Francisco Bay Area, LinkedIn) as an example of (title, region, company) cohorts.

### 2.1 Modeling Challenges

*Company similarity mining and modeling with peer company groups*: The existing salary modeling system had two key limitations for computing company-level insights: (1) There was no intermediary level in the hierarchy between (title, region) and (title, region, company). Consequently, the prior estimates for small-sized cohorts at the company-level were typically derived from the corresponding (title, region) cohort. This approach has the limitation that for the same job title and location, different companies may pay differently and hence using the same ancestral cohort as the prior is not desirable. (2) Insights were generated only for cohorts with at least a few member submissions. In particular, insights for (title, region, company) cohorts with no data at all could not be obtained.

A key modeling challenge to address these limitations is to determine similarity between companies and compute a peer company group for each company. Given peer company groups, the next challenge is to address the above limitations through better statistical smoothing for company-level cohorts with some submissions and robust inference for such cohorts with no data. The proposed model should be able to combine information from member submitted compensation entries (where available) and the computed company similarities, with appropriate weighting to achieve sufficient product coverage and reliable, robust insights.

*Evaluation*: The sensitive and non-public nature of compensation data poses unique evaluation challenges, which are not present in other user-facing products such as recommendation systems. We cannot perform online A/B testing to evaluate different models or other user feedback dependent evaluation approaches, since users may not be aware of the correct compensation range. An additional challenge is the dearth of reliable ground truth compensation datasets. Even when available (e.g., BLS OES dataset [1]), our experience suggests that mapping them to LinkedIn's taxonomy is noisy.

## 3. MODEL AND ALGORITHMS

### 3.1 Peer Company Mining

We next present our approach for computing peer company groups, which could serve as an intermediary level between (title, region) and (title, region, company) in the hierarchy, and thereby help obtain better prior compensation estimates for a company-level cohort. We consider two companies to be similar if employees are very likely to move from one company to the other and vice versa. We assume that in the absence of any other information, fixing the title and the location, for a given company, the set of companies whose employees have transitioned to and from this company could provide a reasonable guidance on the compensation at this company. This assumption is based on the observation that job changes typically result in same or higher pay [22].

We first formally define the notion of peer score between two companies. We then present an algorithm (*Company2vec*) for learning company embeddings from the LinkedIn members' company transition data, which uses techniques such as negative sampling and stochastic gradient descent to map each company to its latent representations. Since our definition of peer companies considers the directed transitions in both directions, yet a company may act differently as a transition origin or destination, origin and destination embeddings are distinguished and modeled separately for each company. From these embeddings, we then compute the peer scores, and also obtain the peer company group for each company as the set of most similar companies, which is used as part of the Bayesian statistical model for smoothing and inferring company-level insights (§3.2). Note that we analyze the ($company_1 \rightarrow company_2$) transitions instead of (($title_1, region_1, company_1$) $\rightarrow$ ($title_2, region_2, company_2$)) transitions due to lack of enough support at finer granularities, and also for simplicity of modeling.

#### 3.1.1 Definitions (Peer Company & Peer Score)

Two companies $u$ and $v$ are *peer companies* if company $v$ is among top choices for employees in company $u$ to transition to and vice versa, and this similarity between companies $u$ and $v$ is measured via *peer score* defined as

$$ps(u,v) := \frac{\mathbb{P}(c_1 = v \mid c_0 = u)}{\max_w \mathbb{P}(c_1 = w \mid c_0 = u)} \cdot \frac{\mathbb{P}(c_1 = u \mid c_0 = v)}{\max_w \mathbb{P}(c_1 = w \mid c_0 = v)}, \quad (1)$$

where $c_0$ denotes the company prior to the transition (origin), $c_1$ denotes the company after the transition (destination), and $\mathbb{P}(c_1 = v \mid c_0 = u)$ is thus the probability of a member transitioning to company $v$ conditioned on the current company being $u$. Peer score, $ps(u,v)$ has a range of $[0,1]$, and reaches its maximum of 1 when companies $u$ and $v$ are each other's top transition choice, i.e.,

$$\begin{aligned} v &= \underset{w}{\operatorname{argmax}}\, \mathbb{P}(c_1 = w \mid c_0 = u), \text{ and} \\ u &= \underset{w}{\operatorname{argmax}}\, \mathbb{P}(c_1 = w \mid c_0 = v). \end{aligned} \quad (2)$$

Without loss of generality, $\mathbb{P}(c_1 = v \mid c_0 = u)$ is denoted as $\mathbb{P}(v \mid u)$ in the rest of this paper.

#### 3.1.2 Member Company Transitions

At LinkedIn, we have rich company transition data from member profiles. As part of LinkedIn profile, *Experience* collects users' work experiences, and each piece of experience contains information such as company, position, and start and end times. For each member, we arrange work experiences into a list of company transitions in time order, and use the transitions as positive samples in the training data. For example, if a member lists consecutive work experiences in Companies $A$, $B$, $C$ in time order with no overlap, then Company $A$ to $B$ transition and $B$ to $C$ transition are marked as positive in training.

#### 3.1.3 Negative Sampling

We apply negative sampling [21] to approximately estimate the transition probability and calculate the peer score. The main idea is to map each company $u$ to its embeddings: i) $\boldsymbol{\phi}_u$ in the latent transition origin space $\boldsymbol{\Phi} \subset \mathbb{R}^m$, and ii) $\boldsymbol{\psi}_u$ in the latent transition destination space $\boldsymbol{\Psi} \subset \mathbb{R}^m$. For each company $u$, we randomly draw $K$ companies not sharing any transition with company $u$ as negative samples, and calculate the transition probability as

$$\mathbb{P}(v \mid u) = \sigma(\boldsymbol{\phi}_u^T \boldsymbol{\psi}_v) \prod_{k=1}^{K} \sigma(-\boldsymbol{\phi}_u^T \boldsymbol{\psi}_{w_k(u)}), \quad (3)$$

where $\sigma(x) = 1/(1 + \exp(-x))$ is the sigmoid function, and $\mathcal{N}_u := \{w_1(u), w_2(u), \cdots, w_K(u)\}$ denotes the set of $K$ randomly sampled negative companies of company $u$. Latent space dimension $m$ and negative sample size $K$ are two parameters to be chosen empirically based on the data size.

With the embeddings, the peer score can be approximately computed by randomly marginalizing out the denominator in Equation (1) [21] as:

$$ps(u,v) \approx \frac{\sigma(\boldsymbol{\phi}_u^T \boldsymbol{\psi}_v)}{\max_w \sigma(\boldsymbol{\phi}_u^T \boldsymbol{\psi}_w)} \cdot \frac{\sigma(\boldsymbol{\phi}_v^T \boldsymbol{\psi}_u)}{\max_w \sigma(\boldsymbol{\phi}_v^T \boldsymbol{\psi}_w)}. \quad (4)$$

#### 3.1.4 Update Procedure

The problem can be then interpreted as an optimization problem of learning the set of latent transition origin embeddings $\boldsymbol{\Phi}_\mathcal{C} = \{\boldsymbol{\phi}_u : u \in \mathcal{C}\}$ and the set of destination embeddings $\boldsymbol{\Psi}_\mathcal{C} = \{\boldsymbol{\psi}_u : u \in \mathcal{C}\}$ for all companies $\mathcal{C}$ with objective function as the log likelihood of the set of all pairs of transitions $\mathcal{T}$ (assuming independence).

$$\arg\max_{\boldsymbol{\phi}_\mathcal{C}, \boldsymbol{\psi}_\mathcal{C}} \sum_{(u,v) \in \mathcal{T}} \sum_{\substack{z \in \\ \{v\} \cup \mathcal{N}_u}} \mathbb{1}_{\{z=v\}} \cdot \log \left[\sigma(\boldsymbol{\phi}_u^T \boldsymbol{\psi}_z)\right]$$
$$+ (1 - \mathbb{1}_{\{z=v\}}) \cdot \log \left[1 - \sigma(\boldsymbol{\phi}_u^T \boldsymbol{\psi}_z)\right]\}. \quad (5)$$

This optimization problem can be solved by stochastic gradient decent (SGD), and iterating between updating origin and destination embeddings until convergence. With a chosen learning rate $\eta$ in SGD, the pseudocode of solving for origin embedding $\boldsymbol{\phi}_u$ for each positive ordered transition pair $(u, v)$ given destination embeddings of company $v$ and all negative samples $\mathcal{N}_u$ of origin company $u$ as input is shown as an example in Algorithm 1. The destination embedding is updated analogously given origin embeddings.

#### 3.1.5 Generate Peer Company Group

The peer score can be calculated as in Equation (4) with the learned origin and destination embeddings $\boldsymbol{\Phi}_\mathcal{C}$ and $\boldsymbol{\Psi}_\mathcal{C}$ produced by SGD. With peer score as similarity measure, for each company, we then generate a list of its peer companies ranked by peer scores in descending order and filtered with a minimum score value.

### 3.2 Bayesian Model for Inferring Insights for Empty Title-Region-Company Cohorts

We next present a flexible Bayesian statistical model for predicting the compensation range for empty (title, region, company) cohorts, utilizing both the company related information present in

**Algorithm 1** Learning origin embedding in *Company2vec*

---
**input**: $\{\boldsymbol{\psi}_z : z \in \{v\} \cup \mathcal{N}_u\}$, i.e., destination embeddings of the positive destination company $v$ and all negative samples in $\mathcal{N}_u$.
**output**: $\boldsymbol{\phi}_u$, i.e., origin embedding of company $u$.
**procedure** UPDATE($\boldsymbol{\phi}_u$)
    $\boldsymbol{e} \leftarrow \boldsymbol{0}$                                                  ▷ Initiate $\boldsymbol{e}$ to be $\boldsymbol{0}$
    **for** $z \in \{v\} \cup \mathcal{N}_u$ **do**
        $g \leftarrow \eta \cdot [\mathbb{1}_{\{z=v\}} - \sigma(\boldsymbol{\phi}_u^T \boldsymbol{\psi}_z)]$
        $\boldsymbol{e} \leftarrow \boldsymbol{e} + g \cdot \boldsymbol{\psi}_z$
        $\boldsymbol{\psi}_z \leftarrow \boldsymbol{\psi}_z + g \cdot \boldsymbol{\phi}_u$
        $\boldsymbol{\phi}_u \leftarrow \boldsymbol{\phi}_u + \boldsymbol{e}$
    **end for**
**end procedure**

---

member submitted compensation data and company similarities mined from LinkedIn members' company transition data using *Company2vec* technique (§3.1). The main idea is to decouple the submitted (title, region, company) compensation data into two components: i) (title, region)-wise compensation and ii) company-wise compensation adjustments, study them separately, and then integrate the inferences from both models together to obtain predictions for (title, region, company) compensation. There is a lot of heterogeneity both in compensation for the same title for different regions (e.g., Software Engineers in San Francisco vs. New York), and in compensation for the same region for different titles (e.g., Software Engineers vs. Nurses in New York). Therefore, instead of using title only or region only component, we choose (title, region) as an integrated component in decoupling. The (title, region) component leverages regression model based prediction approach from [18], while the company component is modeled via a Bayesian model where a company is smoothed with peer company compensation data if there are enough submissions to its peer companies regardless of which (title, region) the submissions are from, and smoothed by global information of all submitted compensation data otherwise. We then recouple results from both (title, region) and company components to generate predictions for (title, region, company) compensation insights using statistical tools.

### 3.2.1 Information Decoupling & Recoupling

The compensation data for a specific (title, region, company) is assumed to follow log normal distribution, which is validated in previous work [18], and its logarithm is denoted as $y_{(t,r,c)}$. The data is decoupled into two parts as i) information explained by $(title, region)$ and ii) company residual (both in the logarithmic space):

$$y_{(t,r,c)} = \mu_{(t,r)} + \epsilon_{(t,r,c)}, \quad (6)$$

where $\mu_{(t,r)}$ is the (title, region) specific mean, and $\epsilon_{(t,r,c)}$ is the residual whose variance is a combination of company adjustment and random noise. We then eliminate the influence of title and region by subtracting the estimated (title, region) mean $\hat{\mu}_{(t,r)}$, and focus on studying company adjustments denoted as

$$\tilde{y}_c := y_{(t,r,c)} - \hat{\mu}_{(t,r)}. \quad (7)$$

After obtaining analysis of both (title, region) and company components, we then recouple the results to get the predicted mean and variance of (title $t$, region $r$, company $c$) compensation as

$$\hat{\mathbb{E}}[y_{(t,r,c)}] = \hat{\mathbb{E}}[\tilde{y}_c | \mathcal{D}_\mathcal{C}] + \hat{\mu}_{(t,r)}, \quad (8)$$
$$\hat{\mathbb{V}}ar[y_{(t,r,c)}] = \hat{\mathbb{V}}ar[\tilde{y}_c | \mathcal{D}_\mathcal{C}] + \hat{\mathbb{V}}ar[\hat{\mu}_{(t,r)}], \quad (9)$$

where $\hat{\mu}_{(t,r)}$ and $\hat{\mathbb{V}}ar[\hat{\mu}_{(t,r)}]$ are (title, region) specific predicted mean and variance from regression in previous work [18]; $\hat{\mathbb{E}}[\tilde{y}_c | \mathcal{D}_\mathcal{C}]$ and $\hat{\mathbb{V}}ar[\tilde{y}_c | \mathcal{D}_\mathcal{C}]$ denote the posterior mean and variance of company adjustment from the Bayesian model, and $\mathcal{D}_\mathcal{C} = \{\tilde{y}_c : c \in \mathcal{C}\}$ is the set of company adjustments for all companies $\mathcal{C}$ used for both calculating priors and Bayesian updating. We assume that $\mathbb{C}ov[\tilde{y}_c, \mu_{(t,r)}] = 0$ (following [18]), that is, the covariance between (title, region) and company components is uncorrelated and thus eliminated from analysis.

### 3.2.2 Bayesian Smoothing

We use a Bayesian model since it provides a flexible structure for incorporating external knowledge in the form of a *prior*. For company $c$ to be studied, let $\mathcal{D}_c = \{\tilde{y}_{c,1}, ..., \tilde{y}_{c,n_c}\}$ denote the set of $n_c$ company adjusted data of company $c$ regardless of the title or region the data belongs to; $pc(c)$ denote its peer company group, which contains a list of companies similar to $c$; and $n_{pc(c)} = \sum_{c' \in pc(c)} n_{c'}$ denote the total number of compensation submission entries for companies in $pc(c)$.

The prior mean and variance of the company component are set to be centered at $(\mu_0, \sigma_0^2)$, which can be obtained from either peer company information or global information. In case of the former, peer company information $(\hat{\mu}_{pc(c)}, \hat{\sigma}_{pc(c)}^2)$ is estimated from all company adjustments of $c$'s peer companies as

$$\hat{\mu}_{pc(c)} = \sum_{c' \in pc(c)} \sum_{i=1:n_{c'}} \tilde{y}_{c',i} / n_{pc(c)}; \quad (10)$$
$$\hat{\sigma}_{pc(c)}^2 = \sum_{c' \in pc(c)} \sum_{i=1:n_{c'}} (\tilde{y}_{c',i} - \hat{\mu}_{pc(c)})^2 / n_{pc(c)}. \quad (11)$$

The prior mean and variance can also be chosen to be global information estimated by all the $n_\mathcal{C} = \sum_{c \in \mathcal{C}} n_c$ compensation submission entries over the set of all companies $\mathcal{C}$ as

$$\hat{\mu}_{all} = \sum_{c' \in \mathcal{C}} \sum_{i=1:n_{c'}} \tilde{y}_{c',i} / n_\mathcal{C}; \quad (12)$$
$$\hat{\sigma}_{all}^2 = \sum_{c' \in \mathcal{C}} \sum_{i=1:n_{c'}} (\tilde{y}_{c',i} - \hat{\mu}_{all})^2 / n_\mathcal{C}. \quad (13)$$

In our application, a company's prior is chosen to be peer company information when the size of its peer company group, $n_{pc(c)}$, is no smaller than a certain threshold $n_\tau$, and centered at global information otherwise. That is,

$$(\mu_0, \sigma_0^2) = \begin{cases} (\hat{\mu}_{pc(c)}, \hat{\sigma}_{pc(c)}^2) & \text{if } n_{pc(c)} \geq n_\tau, \\ (\hat{\mu}_{all}, \hat{\sigma}_{all}^2) & \text{otherwise.} \end{cases} \quad (14)$$

All data in $\mathcal{D}_c$ can be modeled as normal distribution with a conjugate normal-inverse-Gamma prior as

$$\text{Model}: \tilde{y}_{c,i} \sim N(\mu_c, \tau^2) \text{ for } i = 1, \cdots, n_c, \quad (15)$$
$$\text{Priors}: \mu_c | \tau^2 \sim N(\mu_0, \tau^2 / n_0), \quad (16)$$
$$\tau^{-2} \sim Gamma(\eta/\sigma_0^2, \eta), \quad (17)$$

where $n_0 = m/\delta$; $m$ represents the amount of data used to derive the prior, i.e., $m = n_{pc(c)}$ when using peer company information as prior, and $m = n_\mathcal{C}$ when using global information as prior; $\delta$ and $\eta$ are two smoothing hyperparameters indicating how much information is passed from prior to model, which can be optimized via cross-validation. The smaller $\delta$ and $\eta$ are, the more information is passed from the prior. It should be noted that the prior mean of $\mu_c$ is the same as the external data mean $\mu_0$, while the prior distribution of its precision $\tilde{\tau}$ (that is, $\tau^{-2}$ is also centered at external precision mean $1/\sigma_0^2$.

Denoting the mean of entries in $\mathcal{D}_c$ as $\bar{y}_c$, the posterior can be updated as

$$\mu_c|\tau^2, \mathcal{D}_\mathcal{C} \sim N(\frac{n_c}{n_c + n_0}\bar{y}_c + \frac{n_0}{n_c + n_0}\mu_0, \frac{\tau^2}{n_c + n_0}), \quad (18)$$

$$\tau^{-2}|\mathcal{D}_\mathcal{C} \sim Gamma(\frac{n_c}{2} + \frac{\eta}{\sigma_0^2},$$

$$\eta + \frac{1}{2}\sum_{i=1}^{n_c}(\tilde{y}_{c,i} - \bar{y}_c)^2 + \frac{n_c n_0}{2(n_c + n_0)}(\bar{y}_c - \mu_0)^2). \quad (19)$$

By marginalizing out the mean parameter $\mu_c$ and precision parameter $\tau^{-2}$, the posterior prediction $\tilde{y}_c^*$ for company $c$ can be shown to follow a $t$ distribution [27]:

$$\tilde{y}_c^*|\mathcal{D}_\mathcal{C} \sim t_{df_c}(m_c, s_c), \quad (20)$$

where

$$df_c = n_c + \frac{2\eta}{\sigma_0^2}, \quad (21)$$

$$m_c = \frac{n_c}{n_c + n_0}\bar{y}_c + \frac{n_0}{n_c + n_0}\mu_0, \quad (22)$$

$$s_c = (1 + \frac{1}{n_c + n_0})\frac{\eta + \frac{1}{2}[\sum_{i=1}^{n_c}(\tilde{y}_{c,i} - \bar{y}_c)^2 + \frac{n_c n_0}{n_c + n_0}(\bar{y}_c - \mu_0)^2]}{n_c/2 + \eta/\sigma_0^2}. \quad (23)$$

We note that the posterior mean $m_c$ is a weighted sum of data mean $\bar{y}_c$ and prior mean $\mu_0$, while the posterior variance is a combination of data variance, $\sum_{i=1}^{n_c}(\tilde{y}_{c,i} - \bar{y}_c)^2$ and departure of data mean from prior mean, $(\bar{y}_c - \mu_0)^2$.

### 3.2.3 Update Procedure

To summarize, the update procedure for the Bayesian smoothing algorithm is as follows:

1. Run regression on $y_{(t,r)}$ and get (title, region) specific mean estimate $\hat{\mu}_{(t,r)}$ and variance estimate $\hat{\mathbb{V}}ar[\hat{\mu}_{(t,r)}]$.

2. **Decouple** data to get company residual $\tilde{y}_c$ as in Eqn. (7).

3. Run Bayesian smoothing on $\tilde{y}_c$ to get mean estimate $\hat{\mathbb{E}}[\tilde{y}_c|\mathcal{D}_\mathcal{C}] = m_c$ and variance estimate $\hat{\mathbb{V}}ar[\tilde{y}_c|\mathcal{D}_\mathcal{C}] = s_c$.

   - If $n_{pc(c)} \geq n_\tau$, set prior as $(\mu_0, \sigma_0^2) = (\hat{\mu}_{pc(c)}, \hat{\sigma}_{pc(c)}^2)$ and update posterior as in Eqns. (20) to (23).
   - If $n_{pc(c)} < n_\tau$, set prior as $(\mu_0, \sigma_0^2) = (\hat{\mu}_{all}, \hat{\sigma}_{all}^2)$ and update posterior as in Eqns. (20) to (23).

4. **Recouple** both model results to get compensation insights for (title, region, company) as in Eqns. (8) and (9). These equations provide the predicted mean and variance in the logarithmic space, from which the associated log-normal distribution (as well as the displayed percentiles) can be obtained as in [18].

## 4. SYSTEM DESIGN AND ARCHITECTURE

We describe the overall design and architecture of LinkedIn Salary computation system, with a focus on the generation of peer company groups and the inference of company-level insights (see Figure 2). As discussed in [18], the LinkedIn Salary modeling system consists of both online and offline components, which we describe here for completeness. The online component is for compensation insight retrieval corresponding to the query from the user facing product, while the offline component is for insight generation, including *Flows Run As Needed* for training of models and knowledge mining, as well as *Hadoop Flow Run Regularly* for periodic generation of compensation insights.

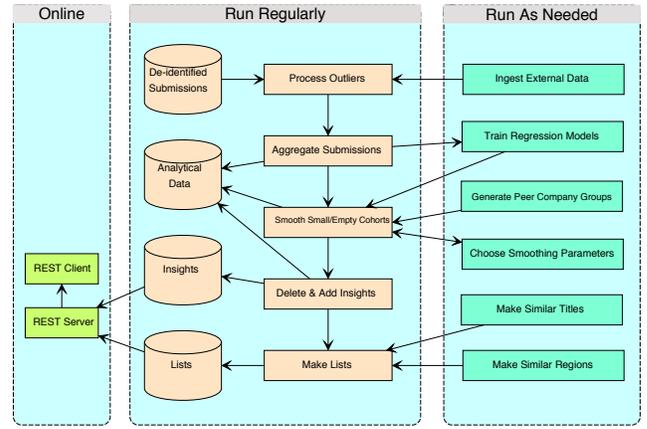

Figure 2: LinkedIn Salary online insight retrieval and offline computing system architecture.

### 4.1 Online System for Insight Retrieval

#### 4.1.1 LinkedIn Salary Platform

The presented compensation insights cover different career dimensions including title, region, industry, company, and years of experience. Upon request by instances of REST Client, compensation insights are provided by the REST Server based on the features of requested cohort. When there are enough submissions for the requested cohort, the insights include the empirical quantiles (10th and 90th percentiles, median) for base salary, bonus, and other compensation types, together with a histogram for base salary. When the data is not sufficient, we report quantiles based on the underlying statistical models to ensure robust insights.

#### 4.1.2 Related Company List

For requests with company as a career dimension, a list of related company compensation insights is also provided to facilitate users to compare compensation with similar companies. For a given company, we obtain this list of companies based on similarity (peer score) (§3.1) and the median compensation.

### 4.2 Offline System for Insight Generation

An offline workflow generates compensation insights by applying statistical and machine learning algorithms to the *De-identified Submissions* data stored on HDFS. The generated insights are then pushed to the *Insights and Lists Voldemort* key-value stores [24], which are probed by the REST Server. The offline workflow is composed of two parts: i) *Run Regularly* for the Hadoop flow that runs more than once a day to update insights with newly collected data, and ii) *Run As Needed* for Hadoop and other flows that run as needed for data mining and model training.

#### 4.2.1 Hadoop Flow Runs Regularly

This Hadoop flow starts with steps, *Process Outliers* and *Aggregate Submissions*, applying sanity check and several statistical methods to detect and remove questionable submissions and then aggregating individual data into cohort-level percentiles and his-tograms. *Smooth Small/Empty Cohorts* uses hierarchical smoothing [18] to obtain robust insights for small-sized cohorts and our proposed Bayesian inference model (§3.2) to infer insights for cohorts with no data. This step consumes the models trained and results mined from *Run As Needed*. Then, *Delete & Add Insights* removes certain insights and adds others from trusted reliable sources, and finally, *Make Lists* generates lists of insights or their keys.

### 4.2.2 Flows Run As Needed

This group consists of components that are mostly independent from each other. *Ingest External Data* maps external data (e.g., BLS OES dataset [1]) to the LinkedIn taxonomy, which is used for outlier detection. *Train Regression Models* trains regression models, *Generate Peer Company Groups* computes company similarity, and *Choose Smoothing Parameters* (see §5.2.4) optimizes tuning parameters. The results of those three components are used for smoothing and prediction in *Smooth Small/Empty Cohorts*. *Make Similar Titles* and *Make Similar Regions* generate a list of similar titles for a given title and a list of similar regions for a given region respectively.

## 5. EXPERIMENTS

We next perform an experimental study of our system. We present an evaluation of our approach for computing peer companies in §5.1, followed by an evaluation of our Bayesian model for accurate and robust estimation of compensation insights for (title, region, company) cohorts in §5.2.

## 5.1 Evaluation of Peer Company Computation

We compare the proposed *Company2vec* approach with the well-known *Word2vec* approach [21] which, in our application context, models the co-occurrence of companies using cosine similarity of their latent embeddings. For each of these two methods, we study the extent to which the results computed by the algorithm agrees with a ground truth dataset. Specifically, we compute the Spearman rank correlation coefficient [23] of the algorithm results with respect to the similar company ranking given in the ground truth dataset. This measure is high when the two ranked lists are mostly in agreement, reaches a maximum of 1 when the two ranked lists are identical, and reaches a minimum of $-1$ when the two ranked lists are in opposite order.

### 5.1.1 Dataset

The training data corresponds to the member's company transition data we obtain from LinkedIn member profiles, where each row denotes the transition of a member from one company to another. If $n_{(u,v)}$ members have moved from company $u$ to company $v$, then there are $n_{(u,v)}$ rows of $(u, v)$ in the dataset. All companies that appear less than 100 times in the training data are discarded to reduce noise, resulting in a dataset with about $190K$ US companies.

The commonly used datasets for testing word embedding algorithms, e.g., *WordSim353* [12] and *SCWS* [17], do not cover company names, and thus cannot be used here. Instead, we use a database of similar companies internally developed (and periodically updated) at LinkedIn as ground truth in the experiment. Each record of this dataset is a company with a ranked list of its similar companies.

### 5.1.2 Comparison Results

Table 1 presents the average Spearman rank correlation score obtained by *Company2vec* and *Word2vec* at various parameter configurations. NEG-$K$ corresponds to applying negative sampling with $K$ negative samples for each company transition pair ($fromCompany$, $toCompany$). The dimension of both transition origin and destination latent spaces is set to 20. We can see that each algorithm has similar performance for negative sample size parameter in the range of *5-20*, and at all three tested parameter configurations, *Company2vec* outperforms *Word2vec* on the similar company ranking task with respect to the chosen ground truth data.

|        | *Company2vec* | *Word2vec* |
|--------|---------------|------------|
| NEG-5  | **0.898**     | 0.530      |
| NEG-10 | **0.896**     | 0.532      |
| NEG-20 | **0.894**     | 0.534      |

Table 1: Performance comparison of *Company2vec* and *Word2vec* in terms of the average Spearman rank correlation score, for different negative sample size parameter choices.

Table 2 presents some sample outputs of peer companies obtained by *Company2vec* and *Word2vec*. We can see that *Word2vec* output exhibits flaws in certain cases (highlighted in bold in Table 2). In these cases, the peer companies computed by *Word2vec* are at very different scale (in terms of company size) compared to the base company. With Goldman Sachs as an example, there are well known peer companies such as J. P. Morgan and Barclays, yet the peer company results computed by *Word2vec* for Goldman Sachs are all small investment companies rather than large well-known ones. A possible explanation could be that the cosine similarity measure in *Word2vec* does not consider the combined bidirectional transition probabilities between companies. The peer score measure (see Equation 4) in *Company2vec*, however, balances the transition probabilities for both directions, such that when the transition probability is high in one direction yet low in the other direction, the peer score is still low.

| Model | Shell | CNN | Goldman Sachs |
|-------|-------|-----|---------------|
|       | **Shelly Global Solutions** | CNN Newsource | **Hedge Fund** |
|       | E.ON UK | **CTV News** | **Nikko** |
| *Word2vec* | Shell Exploration & Production | NBC News | **LCH** |
|       | ExxonMobil | **Black Family Channel** | **Wooris** |
|       | **Equiva** | ABC news | **Millburn** |
|       | Motiva | ABC News | Barclays |
|       | BP | FOX News | AQR Capital |
| *Company2vec* | Shell Exploration & Production | CNN Newsource | JP Morgan |
|       | ExxonMobil | MSNBC | Credit Suisse |
|       | Maersk Oil | CBS News | Evercore Partners |

Table 2: Comparison of the closest companies given by *Word2vec* model and the proposed *Company2vec* model (Shell, CNN, and Goldman Sachs are the input company examples). The two models are trained on member transition data of over $190K$ companies.

## 5.2 Evaluation of Bayesian Model

### 5.2.1 Experiment Setup

We performed our experiments on nearly two years of de-identified compensation data collected across three countries (US, Canada, and UK) and of two compensation types (base salary and total salary). Due to the privacy requirements, we have access only to cohort level de-identified compensation submissions (e.g., base salaries for Software Engineers at LinkedIn in San Francisco Bay Area).

The proposed Bayesian smoothing model is applied separately to all combinations of the three nations and two compensation types, and the performance of base salary data in the US is summarized in this section as an example. We apply data sanity check and outlier detection, then randomly draw 80% of the (title, region, company) cohorts as training set, and use the remaining 20% as testing set.

The model validation is done from two perspectives: i) choice of information size of company, peer company, title, and region to ensure enough data for borrowing information across cohorts (§5.2.3); and ii) hyperparameter tuning (§5.2.4). As a result, a common set of thresholds on company, peer company, title and region size as well as (country, compensation type) specific sets of optimal hyperparameters is chosen to perform robust LinkedIn salary predictions for empty cohorts. Model performance is then evaluated with respect to various model robustness measures, accuracy measures, range measure, and business metrics, and summarized in §5.2.5.

### 5.2.2 Evaluation Metrics

The model performance is evaluated using six measures: i) 80% credible interval (CI) coverage; ii) lower 10% interval coverage; iii) upper 10% interval coverage; iv) root-mean-square error (RMSE); v) negative log likelihood; and vi) range statistic. Though each measure by itself cannot give a comprehensive evaluation of the model, all measures together can evaluate the model from different perspectives including model robustness, prediction accuracy, and whether the range shown to users looks reasonable.

The first three metrics reflect the model's robustness by checking coverage from several angles. Assuming that compensation follows a log-normal distribution, the model represents the predicted distribution for each cohort as a parametrized log-normal distribution. We can then measure model robustness by checking the proportion of testing data falling into a specified range of the predicted distribution based on training data, which we call the interval coverage. Among all ranges, 80% CI (from 10th percentile to 90th percentile), lower 10% interval (less than 10th percentile), and upper 10% interval (larger than 90th percentile) are the three most important ones and hence the ones used in evaluation, since we report the 10th and 90th percentiles to users in LinkedIn Salary product. For these three measures, the closer the values are to 80%, 10%, and 10% respectively, the better the model performs.

Root-mean-square error (RMSE) and (negative log) likelihood measure the model's prediction accuracy. RMSE measures the average departure of data from the predicted mean, while likelihood measures the probability of the test data matching the predicted distribution. More precisely, likelihood is defined as the (combined) probability of observing the entries in each cohort in the testing set (assuming independence across entries) according to the corresponding log-normal distribution predicted by the model.

Range statistic is defined as the ratio of the difference between the predicted 90th and 10th percentiles to the predicted mean (i.e., (90th percentile − 10th percentile)/mean), and thus measures how tight the reported range is relative to the reported mean. A tight range statistic is preferred in the product since a very wide range relative to the mean (e.g., ($10K, $1M$) with a mean of $50K$) does not provide any useful value to the users. The smaller the range statistic, the more informative the salary insights are for the users. For example, although ($20K, $40K$) and ($200K, $220K$) both have a range of $20K$, the latter intuitively conveys more information to the users as it has a smaller range relative to the mean.

We also report the effect of our model on two business metrics, which correspond to product coverage measured in terms of the number of (title, region, company) cohorts and the corresponding number of monthly active users (MAU) for which/whom we have salary insights respectively. Larger these metrics, the more viable and complete the product is, thereby providing an enhanced user experience. Thus, improving these metrics is a key business objective.

### 5.2.3 Threshold on Information Size

In order to provide robust inferred estimates for (title, region, company) cohorts with no submissions using the proposed Bayesian smoothing model, it is important to have sufficient title, region, company, and peer company information marginally for the inferred cohorts. Therefore, minimum thresholds on marginal counts are chosen for different information sources to ensure that the model has enough information to generate accurate and robust results. We discuss the procedure for choosing the minimum threshold on the combined number of submissions needed for a company (hereafter called 'company submission count threshold'), as an example.

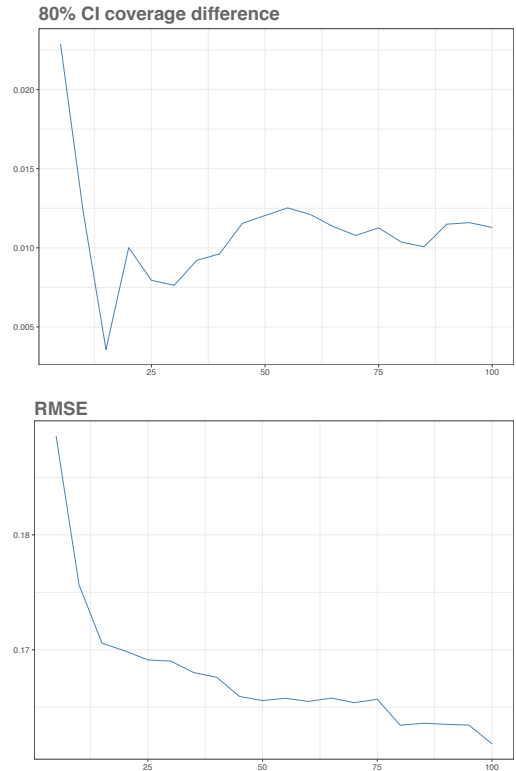

Figure 3: 80% CI coverage difference (upper) and RMSE (lower), plotted against company submission count threshold (for base salary in US).

We evaluate the model performance for different choices of company submission count with respect to all the measures described in §5.2.2. Figure 3 shows the plot of 80% CI coverage difference (defined as |80% CI coverage −0.8|) and RMSE against company submission count threshold. We notice that as the threshold is increased to 25, the model performance metrics become stable after a significant improvement, while there is still a large number of (title, region, company) cohorts that the proposed model can provide insights into ($1203K$ with properly chosen thresholds as stated in §5.2.5). We observed a similar trend for other metrics as well. Therefore we chose 25 as the company submission count threshold.

By performing similar validation on the minimum marginal counts needed for peer company group, title, and region information, we chose the following thresholds: 50 for peer company group, 25 for title, and 40 for region.

### 5.2.4 Optimization of Statistical Smoothing Parameters

The smoothing hyperparameters $\delta$ and $\eta$ are chosen via cross validation within the ranges, $\delta \in \{5r | r \in [1, 20]\}$ and $\eta \in \{0.01 \cdot 2^r | r \in [0, 11]\}$. The two smoothing parameters indicate how much information each company borrows from its peer companies. The smaller the two parameters are, the more information the model borrows from peer company groups ($\delta$ influences the mean while $\eta$ mainly influences the variance smoothing).

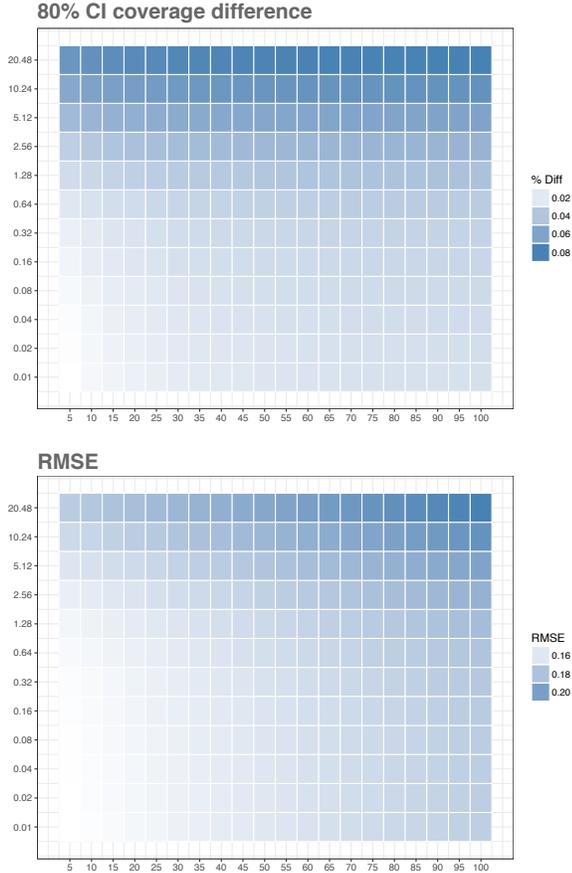

Figure 4: Tuning of hyperparameters, $\delta$ and $\eta$ with base salary in the US as an example (upper: 80% CI coverage difference; lower: RMSE). In both plots, the x-axis and the y-axis represent smoothing parameters, $\delta$ and $\eta$ respectively.

For each set of potential hyperparameter $(\delta, \eta)$ combinations, we evaluate the model performance based on the measures described in §5.2.2, and pick two measures, 80% CI coverage difference and RMSE, as examples in Figure 4. We observed similar pattern for the other measures such as the likelihood. The optimal hyperparameters are chosen to be $\delta = 5$ and $\eta = 0.01$, implying that the peer company information plays an important role in modeling company salary. We performed similar parameter tuning separately on all combinations of countries (US, Canada, UK) and compensation types (base salary, total salary), and observed similar results suggesting that the model relies heavily on the information borrowed from peer companies in each case.

### 5.2.5 Model Performance & Evaluation

We next compare the salary insights inferred by our model for (title, region, company) cohorts in the testing set, with the corresponding insights computed empirically and using the Bayesian hierarchical smoothing model from [18] respectively. Note that these two baseline models make use of the compensation submissions obtained from members for such cohorts in the testing set, while our model does not have access to these submissions and instead performs inference based on the training set. In spite of this access restriction, we show that our model outperforms the baseline approaches. We compare the models using the three interval coverage metrics (§5.2.2), and present the results in Table 3, wherein the ideal values of these metrics are also shown. We observe that our approach significantly improves over the empirical estimates, and also outperforms the Bayesian hierarchical smoothing model [18].

|  | Lower 10% interval coverage | 80% CI coverage (10th to 90th percentile) | Upper 10% interval coverage |
|---|---|---|---|
| Ideal | 10% | 80% | 10% |
| Empirical | 28% | 44% | 28% |
| Previous work [18] | 6% | 87% | 7% |
| Our approach | **9%** | **81%** | **10%** |

Table 3: Comparison of empirical estimate, previous work [18], and our model with respect to the interval coverage metrics (for base salary in the US, as an example).

Moreover, by choosing submission count thresholds of 25 for company, 50 for peer company, 25 for title, and 40 for region to prune cohorts with insufficient marginal information, our model provides compensation insights for $1169K$ additional (title, region, company) cohorts with no data. Together with the $34K$ cohorts with data, we are now able to provide salary insights for $1203K$ cohorts, corresponding to 4.9 times as many monthly active users (summarized in Table 4). Figure 5 shows an example of the predicted salary ranges, which have been deployed as part of LinkedIn Salary product.

|  | Number of covered (title, region, company) cohorts | Relative number of covered monthly active users |
|---|---|---|
| Previous work [18] | 34K | 1x |
| Our approach | **1203K** | **4.9x** |

Table 4: Comparison of product coverage in terms of the number of covered (title, region, company) cohorts and monthly active users.

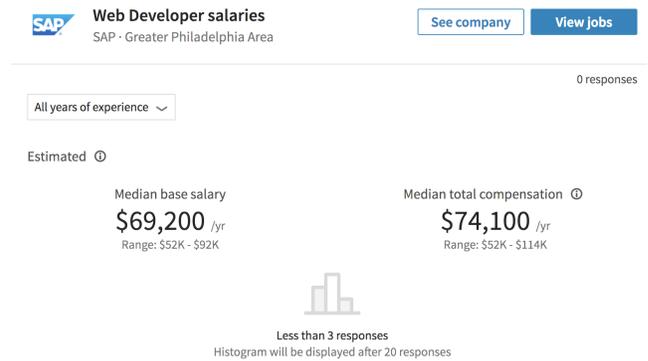

Figure 5: LinkedIn Salary insight page, displaying the predicted salary ranges for Web Developers at SAP in Greater Philadelphia Area.

## 6. LESSONS LEARNED IN PRACTICE

We next present the challenges encountered and the lessons learned through the production deployment of our computation system as part of the LinkedIn Salary platform for more than six months.

### 6.1 Similarity Score for Peer Company Modeling

We first explored using *Word2vec* based similarity measure (cosine similarity) between two companies for peer company modeling. However, we noticed that this measure does not differentiate between moving from one company to another and moving in the opposite direction, and further does not model the combined transition probability in both directions. Hence, instead of adopting existing similarity measures, we introduced the new notion of *peer score* (see Equation 4), which is specifically design for our application scenario. For each pair of companies, the peer score jointly models both transition directions, and uses appropriate normalization to eliminate influence from the scale of a company.

### 6.2 Prior Robustness in Bayesian Smoothing

When applying Bayesian smoothing on a low-support (title, region, company) cohort, the posterior largely depends on the prior since there are very few entries in the cohort. Thus, the prior computed from the corresponding peer company group cohort plays a key role, and it is important to make sure that this prior is robust. In practice, we observed data sparsity for several (title, region, peer company group) cohorts, and decided to apply an additional layer of smoothing in such cases. For such cohorts, we smoothed with the global mean and variance from the corresponding (title, region) cohort.

### 6.3 Filtering by LinkedIn Member Data

Although we can predict the compensation range for a large number of (title, region, company) cohorts by taking the cross product between all (title, region) cohorts and all company components, many of these combinations may not even exist. For example, in our initial results, our model had predicted the compensation range for software engineer positions at LinkedIn in many regions where LinkedIn did not have any presence, as a result of which the corresponding insight webpages were generated as part of the product. Since such cohorts do not provide meaningful value to LinkedIn members and can be thought of as adding noise to the ecosystem, we decided to keep only those cohorts that map to a sufficient number of LinkedIn members. We experimented with different thresholds on the minimum number of members needed for a cohort to be considered valid, and computed the product coverage for each threshold (see Table 5). We chose 2 as the threshold as this choice helps to eliminate spurious cohorts while retaining a significant increase in the product coverage. Note that we did not choose 1 as the threshold as we came across several instances where the title entered by just one member did not match the actual title in the company in the region.

## 7. RELATED WORK

*Salary Information Products*: There are several commercial services offering compensation information. For example, the US Bureau of Labor Statistics [5] publishes a variety of statistics on pay and benefits based on the comprehensive data the government has collected. Glassdoor [4] offers a comparable service, while PayScale [6] sells compensation information to companies.

*Company Embeddings for Peer Company*: Compared with traditional approaches of calculating company similarity based on company features such as company keywords, company sector, and company size [25], our approach learns company embeddings by leveraging LinkedIn's member transition graph. Our *Company2vec* approach is inspired by the *Word2vec* method based on skip-gram with negative sampling [21]. The *Company2vec* approach takes our application setting into account, and defines the new notion of *peer score* instead of using cosine similarity of embeddings as in many applications. Other approaches for composing diverse vector representations into hybrid representations have also been explored [13]. Moreover, the learned company embeddings can be used in other applications, for instance, as features for wide and deep recommender systems [11] and for query expansion [20]. Dynamic change of the transition network [10, 9] could also be of interest for future studies in salary trend analysis.

| Threshold | Number of covered (title, region, company) cohorts | Relative number of covered monthly active users |
|---|---|---|
| 1 | 2761K | 6.9x |
| **2** | **1203K** | **4.9x** |
| 3 | 750K | 4.1x |
| 4 | 535K | 3.6x |
| 5 | 409K | 3.3x |

Table 5: Comparison of the product coverage for different thresholds on the minimum number of members needed for a cohort to be considered valid.

*Statistical Smoothing*: Compared with frequentist approaches, Bayesian statistics provides a flexible structure for incorporating external information [14, 16], with the peer company group information in our application being a good example. The idea of Bayesian smoothing is inspired by the smoothing of sparse events (e.g., CTR) in the context of computational advertising [7, 28], wherein a hierarchy is used so that the combination of an ad and a publisher with very little data can borrow strength from its ancestor nodes. Our model is quite different, both in terms of the application context and the modeling/distributional assumptions.

## 8. CONCLUSIONS AND FUTURE WORK

We studied the problem of reliably inferring compensation insights at the company level, that is, *predicting insights for (title, region, company) cohorts with no member-submitted data at all*. We utilized the rich LinkedIn Economic Graph data to generate a novel, semantic representation (embedding) of companies. We proposed (*Company2vec*), an algorithm for learning company embeddings from the LinkedIn members' company transition data, computed pairwise similarity values between companies based on these embeddings, and then obtained a peer company group for each company as the set of most similar companies. Finally, we computed insights at the company level using a Bayesian statistical model, which made use of company similarities and combined the estimates for (title, region) component and company residual. We demonstrated significant increase in key business metrics such as the number of cohorts with compensation insights and the product coverage in terms of the monthly active users, thereby ensuring the viability of the company-level insight pages in the LinkedIn Salary product. For example, our modeling techniques were responsible for 35 fold increase in the number of (title, region, company) cohorts with insights in the U.S., corresponding to a coverage for 4.9 times as many monthly active users, while slightly improving the quality of the generated insights. We performed extensive experiments with de-identified data collected from millions of LinkedIn members, and also presented the lessons learned from more than six months of production deployment.

Given the desire to improve the quality as well as the coverage of compensation insights towards benefitting more LinkedIn members, we have been pursuing several modeling and engineering directions to extend this work. We plan to develop tools to collect inputs and feedback from recruiters, who typically have better knowledge of the compensation range for their function(s) and industry(ies). Such inputs could be useful for diagnosing cohorts with incorrectly predicted compensation ranges, and potentially correcting them. We would also like to incorporate richer features such as years of experience and skills as part of the prediction model, and detect and correct sample selection bias, response bias, and other biases using statistical techniques. Moreover, as compensation can change over time due to supply/demand changes, inflation, and other economic factors, we would like to take time into consideration when computing salary insights and explore approaches such as discounting and/or appropriately scaling old salary submissions, and building time series models. Finally, we would like to provide personalized compensation estimates for LinkedIn members, by taking into consideration of each member's work experience, education background, skills, and other relevant attributes.

## 9. ACKNOWLEDGMENTS

The authors would like to thank all other members of LinkedIn Salary team for their collaboration for deploying our system as part of the product, and Stuart Ambler, Keren Baruch, Kinjal Basu, Rupesh Gupta, Santosh Kumar Kancha, Myunghwan Kim, Ram Swaminathan, and Ganesh Venkataraman for insightful feedback and discussions.